%% file: tr.tex
\documentclass[12pt]{article}
\usepackage{fullpage}
\usepackage{times}

\ifx\pdftexversion\undefined
  \usepackage[dvips]{graphics}
\else
  \usepackage[pdftex]{graphics}
\fi

\usepackage{amsmath}
\usepackage{rotating}
\usepackage{tabularx}
\usepackage[
pageanchor=true,
plainpages=false,
pdfpagelabels, 
bookmarks,
bookmarksnumbered,
pdfborder=0 0 0,
pdfpagemode=UseOutlines]{hyperref}
\usepackage{draftstamp}
\usepackage{subfigure}

\include{mymacros}

\newcommand{\topicbrowser}{{\footnotesize \textsf{TopicViz}}}
\newcommand{\shiftr}{{\footnotesize \textsf{Shiftr}}}
\newcommand{\topic}[1]{\textbf{#1}}
\newcommand{\example}[1]{``#1''}

\title{TopicViz: Semantic Navigation of Document Collections}
\author{
  Jacob Eisenstein, Duen Horng ``Polo'' Chau, Aniket Kittur and Eric Xing\\
  School of Computer Science\\
  Carnegie Mellon University\\
  Pittsburgh, PA 15215 USA}

\begin{document}
\maketitle
\abstract{
  When people explore and manage information, they think in
  terms of \emph{topics} and \emph{themes}. However, the software that
  supports information exploration sees text at only the surface level.
  In this paper we show how topic modeling -- a technique for
  identifying latent themes across large collections of documents -- can
  support semantic exploration.  We present \topicbrowser, an
  interactive environment for information
  exploration.  \topicbrowser~combines traditional search and
  citation-graph functionality with a range of novel interactive
  visualizations, centered around a force-directed layout that links
  documents to the latent themes discovered by the topic model.  We
  describe several use scenarios in which \topicbrowser~supports rapid
  sensemaking on large document collections.
}
\input{intro}

\input{background}
\input{scenario}
\input{system}
\input{summary}

\section*{Acknowledgments}
This work was supported by the following grants: AFOSR FA9550010247,
ONR N0001140910758, NSF OCI-0943148, NSF IIS-0968484, NSF IIS-0713379, NSF CAREER
DBI-0546594, and an Alfred P. Sloan Fellowship.
\bibliographystyle{abbrv}
\bibliography{cites.bib}
\end{document}

%% file: mymacros.tex
\newcommand{\comment}[1]{}
\newcommand{\bigonly}[1]{}

\def\gcode#1#2{
  #1
  \tt
  \newcommand{\cwhile}{{\bf while}}
  \newcommand{\crepeat}{{\bf repeat}}
  \newcommand{\cuntil}{{\bf until}}
  \newcommand{\cassert}{{\bf assert}}
  \newcommand{\cbreak}{{\bf break}}
  \newcommand{\cexit}{{\bf exit}}
  \newcommand{\ccontinue}{{\bf continue}}
  \newcommand{\cfor}{{\bf for}}
  \newcommand{\cforeach}{{\bf foreach}}
  \newcommand{\cdo}{{\bf do}}
  \newcommand{\cdoparallel}{{\bf do\_parallel}}
  \newcommand{\cdoall}{{\bf doall}}
  \newcommand{\cto}{{\bf to}}
  \newcommand{\cby}{{\bf by}}
  \newcommand{\cif}{{\bf if}}
  \newcommand{\cthen}{{\bf then}}
  \newcommand{\celse}{{\bf else}}
  \newcommand{\cend}{{\bf end}}
  \newcommand{\cnot}{{\bf not}}
  \newcommand{\cand}{{\bf and}}
  \newcommand{\cor}{{\bf or}}
  \newcommand{\cdiv}{{\bf div}}
  \newcommand{\cmod}{{\bf mod}}
  \newcommand{\calgorithm}{{\bf algorithm}}
  \newcommand{\cbegin}{{\bf begin}}
  \newcommand{\creturn}{{\bf return}}
  \newcommand{\cset}{{\bf set}}
  \newcommand{\clist}{{\bf list}}
  \newcommand{\cpair}{{\bf pair}}
  \newcommand{\carray}{{\bf array}}
  \newcommand{\cof}{{\bf of}}
  \newcommand{\cinteger}{{\bf integer}}
  \newcommand{\creal}{{\bf real}}
  \newcommand{\cboolean}{{\bf boolean}}
  \newcommand{\cfalse}{{\bf false}}
  \newcommand{\ctrue}{{\bf true}}
  \newcommand{\cenum}{{\bf enum}}
  \newcommand{\cvar}{{\bf var}}
  \newcommand{\cprefetch}{{\bf prefetch}}
  \newcommand{\cmax}{{\bf max}}
  \newcommand{\cmin}{{\bf min}}
  \newcommand{\csizeof}{{\bf sizeof}}
  \begin{tabbing}
    #2\=#2\=#2\=#2\=#2\=#2\=#2\=#2\=#2\=#2\=#2\=#2\=#2\=\kill
    }
\def\endgcode{\rm \end{tabbing}}

\def\figurecode{\begin{gcode}{\small}{\hspace{0.1in}}}
\def\endfigurecode{\end{gcode}}

\def\smallfigurecode{\begin{gcode}{\footnotesize}{\hspace{0.1in}}}
\def\endsmallfigurecode{\end{gcode}}

\def\tinyfigurecode{\begin{gcode}{\scriptsize}{\hspace{0.1in}}}
\def\endtinyfigurecode{\end{gcode}}

\def\textcode{\begin{gcode}{\normalsize}{\hspace{0.15in}}\+\>}
\def\endtextcode{\end{gcode}}

\def\smalltextcode{\begin{gcode}{\small}{\hspace{0.15in}}\+\>}
\def\endsmalltextcode{\end{gcode}}

\def\tinytextcode{\begin{gcode}{\tiny}{\hspace{0.15in}}\+\>}
\def\endtinytextcode{\end{gcode}}

\def\scriptsizetextcode{\begin{gcode}{\scriptsize}{\hspace{0.15in}}\+\>}
\def\endscriptsizetextcode{\end{gcode}}

\def\footnotesizetextcode{\begin{gcode}{\footnotesize}{\hspace{0.15in}}\+\>}
\def\endfootnotesizetextcode{\end{gcode}}

\newlength{\figratio}
\newlength{\acmratio}


%% file: intro.tex
\section{Introduction}
As information repositories continue to expand and diversify, there is
an urgent need for systems that help people explore and make sense of
large document collections.  While researchers in information-seeking
and related areas have developed increasingly effective interaction
techniques for navigating document
collections~\cite{baldonado1997sensemaker,marchionini2006exploratory},
these methods are hampered by a view of language that is generally
restricted to the surface level; such techniques are oblivious to the
semantic meaning behind the text.  Meanwhile, researchers in machine
learning and natural language processing have developed powerful
statistical methods for recovering latent
semantics~\cite{Blei2003Latent}, but the output of these methods is
difficult to present to both domain expert and novice users alike.

\begin{figure*}
\begin{center}
\includegraphics[width=\textwidth]{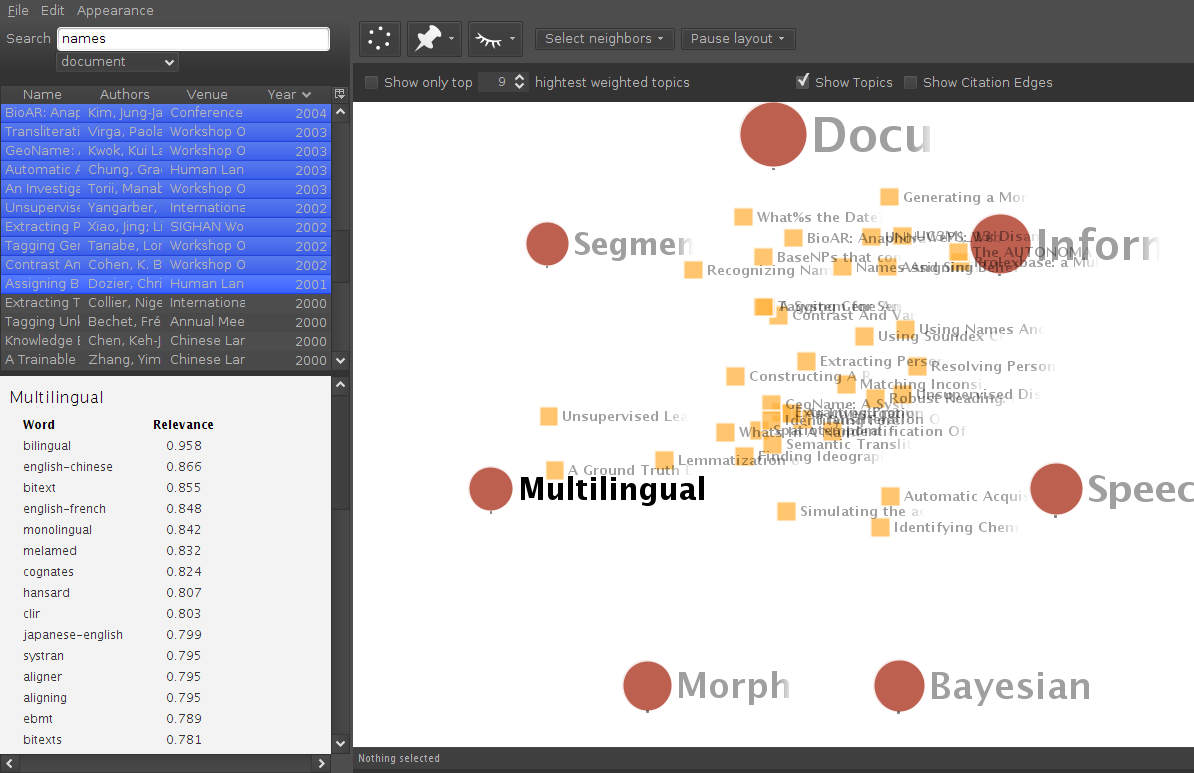}
\caption{The \topicbrowser~environment.  The main panel shows the
initial presentation for a selected set of documents, which are
arranged in a force-directed layout controlled by the seven best-matching
topics.  The upper-left panel shows the search results in list form, 
and the lower-left panel describes the selected topic, ``\topic{multilingual}''.}

\label{f:scenario1-names}
\end{center}
\end{figure*}

In this paper, we introduce~\topicbrowser, a new tool for searching
and navigating large document collections (Figure~\ref{f:scenario1-names}).  
\topicbrowser~infers a set of topics that summarize the latent high-level semantic organization
of a collection, and provides a novel interactive view that exposes
this semantic organization using a force-directed layout.  This layout
permits a range of interactive affordances, allowing users to
gradually refine their understanding of the search results and
citations links, while focusing in on key semantic distinctions of
interest.

The analytic engine of our approach is the topic model -- a powerful
statistical technique for identifying latent themes in a document
collection~\cite{Blei2003Latent}.  Without any annotation, topic
models can extract topics -- sets of semantically-related words -- and
describe each document as a mixture of these topics.  For example, a
given research paper might be characterized as 70\% human-computer
interaction, and 30\% machine learning.\footnote{This distinguishes
  topic models from more coarse-grained techniques that treat each
  document as a member of a single cluster~\cite{cutting1992scatter}.}
Topic models have been successfully applied to a broad range of text,
and the extracted topics have been shown to cohere with readers'
semantic judgments~\cite{Chang2009Reading}.  But while topic models
are often motivated as a technique to support information seeking,
there has been little investigation of how users can understand and
exploit them.

One of the principle strengths of topic models is their flexibility:
topics need not correspond to any predefined taxonomy, but rather
represent the latent structure inherent to the document collection.
However, this means that the content of each topic must somehow be
conveyed to the user.  In topic modeling research, this issue is
almost invariably addressed by showing ranked lists of words and
documents that are closely associated with each topic.  But such lists
have undesirable properties: it is difficult to show more than a few
entries per topic, the meaning of individual terms may be unknown to
non-experts;\footnote{For example, ``muc'', ``muc-6'' and ``muc-7''
  are three of the four most relevant terms for one of the topics
  learned by our model.  These terms are well-known to experts in
  natural language processing (they are the names of shared research
  tasks), but are incomprehensible to an outsider.} in addition, the
numerical scores for each word and topic are hard to interpret.


While hundreds of papers address the mathematical methodology of topic
modeling, relatively few take up the question of how topic models can
support information exploration.  Our approach is distinguished from
prior work in its emphasis on interaction: the user is empowered to
manipulate the visualization by adding, rearranging or removing topics,
and by controlling the set of documents to visualize.  The motivation
for this design stems from our focus on local information exploration:
we aim to provide a deep understanding of a local area of the
information landscape that is relevant to the user's goals, rather
than a surface-level static view of thousands of documents.  We
provide affordances for users to quickly focus in on the topical
distinctions that are relate to their goals, allowing them to
interactively manipulate topics within this space to better understand
document-document, document-topic, and topic-topic relationships.

%% file: background.tex
\section{Background}
\label{s:background}

\paragraph{Topic models of document collections}
A topic model is a hierarchical probabilistic model of document
content~\cite{Blei2003Latent}.  Each topic is a probability
distribution over words, $\beta$; every word in every document is
assumed to be randomly generated from one topic.  In a given document
the proportion of words generated from each topic is given by a latent
vector $\theta_d$.  Thus, the matrix $\boldsymbol{\theta}$ provides a succinct
summary of the semantics of each document.  

Both the topics $\beta$ and the document descriptions $\theta$ can be
obtained through offline statistical
inference~\cite{Blei2003Latent,Griffiths2004Finding}, without any need
for manual annotation.  Thus, topics need not correspond to any
predefined categories; indeed, this is why they are useful for
exploratory analysis.  In the research literature, topic models are often
displayed through textual tables showing the most relevant words and
documents for each topic, as in Figure~\ref{f:topic-text}.



\begin{figure}
\begin{center}
\begin{tabular}{ccc}
``Morphology'' & ``Multilingual'' & ``Parsing'' \\
\hline
morphemes & bilingual & nonterminal \\
morpheme &english-chinese & nonterminals \\
affixes & bitext & adjoining \\
affix & english-french & cfgs \\
kanji & monolingual & cfg\\
endings &melamed & subtree \\
inflections & cognates & non-terminal \\
suffixes & hansard & subtrees \\
inflectional &japanese-english & adjunction \\
katakana  & systran & non-terminals 
\end{tabular}
\end{center}
\caption{A textual display of the top ten automatically-identified keywords 
from three topics obtained from a dataset of research papers on 
computational linguistics.
The topic names were assigned manually.}
\label{f:topic-text}
\end{figure}

In typical scenarios, the number of topics ranges from 10-200, and the
number of documents can range from a few hundred to hundreds of
thousands.  The number of topics can be determined
automatically~\cite{HDP:2006}, or set in advance through interactive
exploration by a domain expert.  Note that \topicbrowser~is not
currently designed to support the user in training new topic models or
exploring alternative topic model parametrizations.  Rather, we target
the case where a topic model is trained in advance, to be used by many
novices who are interested in a given domain, such as legal documents
or research literature.  We consider the problem of supporting end
users to train new topic models to be an important area of future
work.

\paragraph{Visualizing large document collections}
Most prior work on visualizing large document collections can be
divided into two high-level streams: citation graphs and static
projection.  In citation graph approaches, each document is a node, 
and edges are used to represent citation links; clustering is then 
performed over the resulting graph~\cite{Small1999Visualizing,Boyack2005Mapping}.  
The clusters are displayed using techniques such as 
triangulation~\cite{Small1999Visualizing} or force-directed layout~\cite{Boyack2005Mapping}.
Such approaches are well-suited to discover connected disciplines
in science at a high-level, but do not consider the textual content 
of individual documents.

Projection-based methods apply topic
models~\cite{Iwata2008Probabilistic,Herr2009NIH,Maaten2008Visualizing} 
or related techniques like Latent Semantic Analysis~\cite{Landauer2004Paragraph}.
The high-dimensional document descriptions ($\theta_d$ in our notation from
earlier in this section) are then projected into two-dimensional coordinates
for visualization (Landauer et al. user color as an additional
dimension~\cite{Landauer2004Paragraph}).  
A related, recent approach to use visualize topic models is the work of
Liu et al., who emphasize the temporal dimension by placing it on the the
X-axis of a graph that shows the evolution of topic strength and content
over time~\cite{Liu2009Interactive}.

We differ from this prior work in our emphasis on document search and
interactive sensemaking~\cite{Dervin1983Overview,Kuhlthau1991Inside}.
Rather than viewing the entire collection and topic model in a single 
static view, the user manipulates an ever-shifting subset of documents 
and topics.  This approach is driven by the intuition -- dating
back to early work on Scatter/Gather~\cite{cutting1992scatter} -- that only a small
corner of the topic space will be relevant for any given information
search.  We allow documents to be easily added and removed from
the view, either through additional search queries or by exploring citation links;
similarly, topics can be moved and manipulated to reveal subtle semantic distinctions.
The remainder of the paper describes these affordances in greater detail.





%% file: scenario.tex
\section{Scenarios}
The key idea behind \topicbrowser~is to integrate a force-directed
layout for topic models with an integrated environment for expanding
and refining a document list.  As in conventional document search, the
entrance point is the search query; however, rather than simply
listing the search results, they are visualized in an interactive
force-directed layout with a range of affordances.  As
these capabilities are best described by example, this section is
centered around a detailed novice user scenario and two briefer expert
scenarios.\footnote{Video of many of these affordances can be found at
  \url{http://www.cs.cmu.edu/~dchau/topicviz/topicviz.mp4}.} The 
mechanisms underlying \topicbrowser~are described in detail in the 
following section.

\subsection{Novice scenario}
Consider an individual given the task of searching an unfamiliar
research literature, with the goal of identifying whether a particular
technology can be applied to a commercial problem.  In our scenario,
the individual is tasked with determining whether it is possible to
automatically identify names on foreign language websites, using a
collection of 15,032 research papers on computational
linguistics~\cite{Radev2009ACL}.

The user begins by devising a query; with current tools
like Google Scholar and Lexis Nexis, the response to the query would
be an ordered list of results.  Only some of the resulting documents
will be relevant, and almost surely there will be relevant documents
that do not match the query.  The user may then vary the search terms
or navigate the citation links to try to get a complete sense of the
research literature in this unfamiliar area.

\begin{figure}
\begin{center}
\includegraphics[width=.9\textwidth]{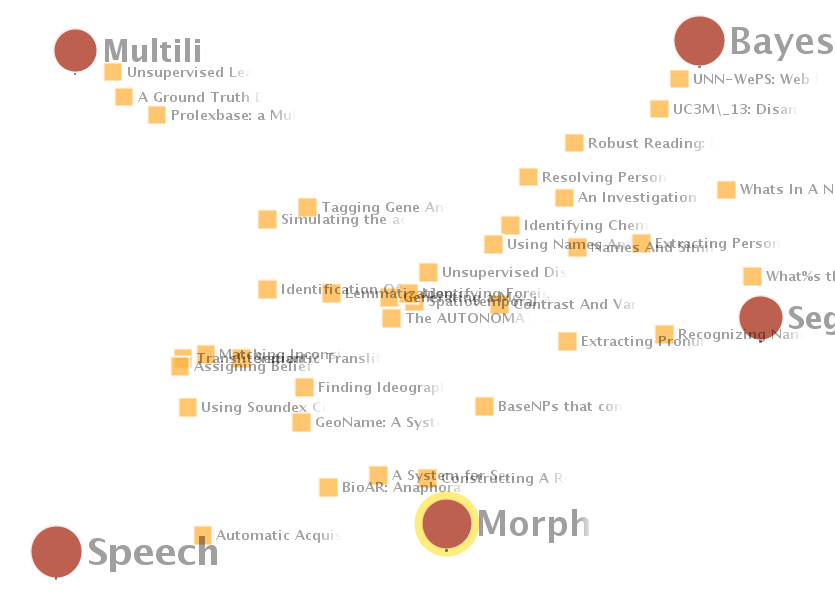}
\caption{Rearranging the topic centers to view topic-topic relationships}
\label{f:scenario1-morphology}
\end{center}
\end{figure}

Now consider the same task, performed with \topicbrowser.  The first
step is the same: the user supplies a search query.  The results are
shown in a list (the top-left part of Figure~\ref{f:scenario1-names}).
The user then drags as many documents as desired into the main area,
which is called the Topic Field: each document is displayed as a node,
and these nodes are surrounded by a ring of topic centers.  The topic
centers are ``pinned,'' while the position of each document is set by
a force-directed layout in which the topics each exert an attractive
force proportional to the document's topical relevance.  Thus,
documents with similar content will be located near each other.

The size of each topic center is determined by its relevance to the
documents in the field, and only the most relevant topics are shown.
The panel on the lower-left shows the most relevant words for each
topic (selected by mouseover).  The user can also see the relevance of
each topic to the documents in the field both statically (by the
document's position) and dynamically (by dragging the topic center
around to see how the document nodes are affected).  The topic names
are specified in advance, either manually by a domain expert, or
through automatic methods~\cite{Mei2007Automatic}; the user is free to
rename topics with more familiar terms.

In our scenario, the user recognizes the topic \topic{multilingual} as
especially relevant to the search -- but other topics like
\topic{morphology} and \topic{Bayesian} are not familiar.  To better
understand if these topics are relevant, the user rearranges the
topics, with \topic{multilingual} in the upper-left corner and the
unfamiliar topics in an arc across the screen
(Figure~\ref{f:scenario1-morphology}).  From this view, the user sees
that \topic{morphology} is related to \topic{multilingual}, as several
documents have strong connections with both topics.  

The user inspects the set of terms associated with the topic
\topic{morphology} (Figure~\ref{f:topic-text}).  While terms like
\example{morpheme} and \example{inflection} are confusing, the user
recognizes the terms \example{affix} and \example{suffix} as referring
to parts of individual words.  Based on this insight, the user renames
the topic from \topic{morphology} to \topic{subwords}.  While this
name is not typically used in the research literature, it helps the
user relate the topic model to her pre-existing ontology.

\begin{figure}
\begin{center}
\includegraphics[width=.8\textwidth]{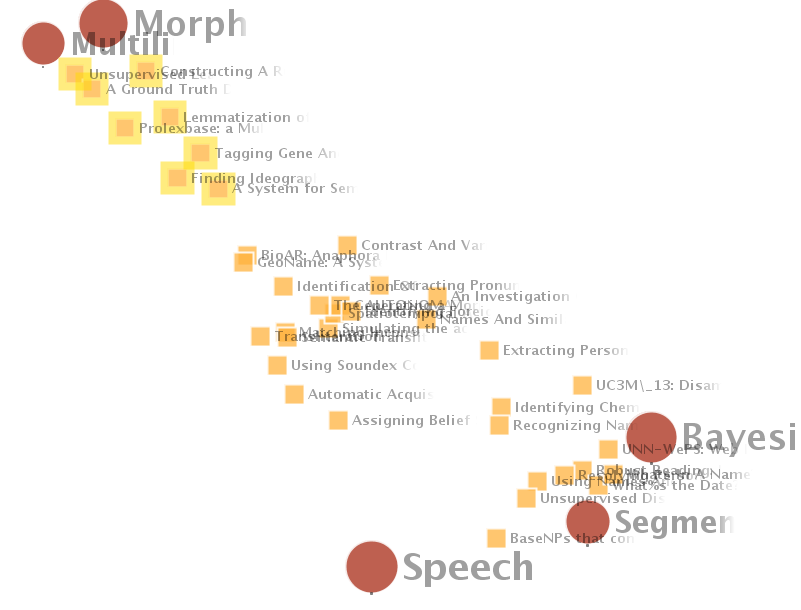}
\caption{By arranging the topic centers into two points, the documents are shown linearly by relevance.}
\label{f:scenario1-spectrum}
\end{center}
\end{figure}

Having identified \topic{morphology} and \topic{multilingual} as key
topics of interest, the user again rearranges the topics, placing the
relevant topics in one corner of the screen and the others in another
corner.  This causes the document nodes to form a line, with location
governed by relevance to the topics of the interest
(Figure~\ref{f:scenario1-spectrum}).  The user now removes documents
that are not close to the desired topics by selecting and deleting
their nodes.

The user has now culled the original list of query hits to a set of
documents that are closely related to multiple topics of interest.
But the coverage of this document set depends on the quality of the
original query.  To make sure that important documents have not been
missed, the user selects a subset of particularly promising documents
and adds documents that cite them.  These new documents may not match
the search query by name, but may still be relevant.  The user can now
investigate the topical characteristics of these new documents and
further refine the search.

Ultimately, the user arrives at a set of documents that reflect the
underlying semantics of the information search.  By investigating the
topic structure, the user has pruned away ``false positives'' that
match the query but are in fact irrelevant; by walking the citation
graph, the user has identified ``false negatives'' that are relevant
but did not match the original query.  Morever, by interatively
exploring the documents, topics, and terms that relate to the initial
query, the user acquires a deeper, structured understanding of the
relevant area of the document collection.  This elucidates the
specific role played by each document in the relevant research
literature, and contextualizes previously unknown themes, such as the
topic \topic{morphology}.  The user is now prepared to summarize the
desired content, having obtained both a comprehensive, high-precision
list of documents and a clearer understanding of this area of research.

\subsection{Expert scenarios}
\begin{figure}
\begin{center}
\includegraphics[width=.8\textwidth]{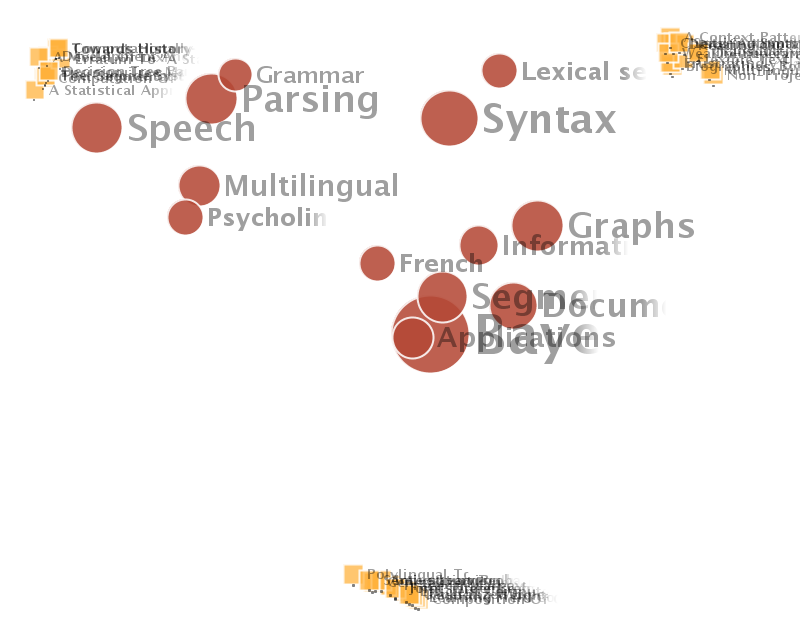}
\caption{To compare topical emphasis of different authors, the expert user
  creates and pins ``piles'' of documents for each author; the
  unpinned topic centers are pulled between them.}
\label{f:authors}
\end{center}
\end{figure}

\paragraph{Determining author expertise}
We briefly consider a scenario involving a user who has more expertise
in the domain of the document collection.  Here, the expert wants to
identify the topical interests of several authors -- perhaps to
distinguish the specific contributions of multiple authors on a single
paper.  To do this, the user searches for papers by each author and
drags them into the field.  However, unlike the previous view, the
\emph{documents} are pinned in place, and the topics float between
them (such non-default behavior can be easily set using the toolbar at
the topic of the window).  The edges in the force-directed layout are bidirectional, and
work identically in this setting; the user need only pin sets of
documents for each author, and then add relevant topics to the view.
Figure~\ref{f:authors} shows such a view for the relationship between
the three authors of a heavily-cited paper in computational
linguistics; this view reveals that the author to the upper-left has
focused more on the \topic{speech} topic; the author to the
upper-right has focused more on \topic{syntax} and \topic{lexical
  semantics}; and the author on the bottom has focused more on the
\topic{Bayesian} and \topic{applications} topics.

\paragraph{Direct manipulation 2D projections}
Finally, we consider a scenario in which an expert user has a detailed
understanding of the topic model, and wants to select a set of
documents that fit a very specific semantic profile.  The novice
scenario explored an affordance in which documents were arranged on a
spectrum between two topics (Figure~\ref{f:scenario1-spectrum}).  In
fact, much more expressive arrangements are possible, yielding a
direct-manipulation inferface for creating two-dimensional projections.  

\begin{figure}
\begin{center}
\includegraphics[width=.9\textwidth]{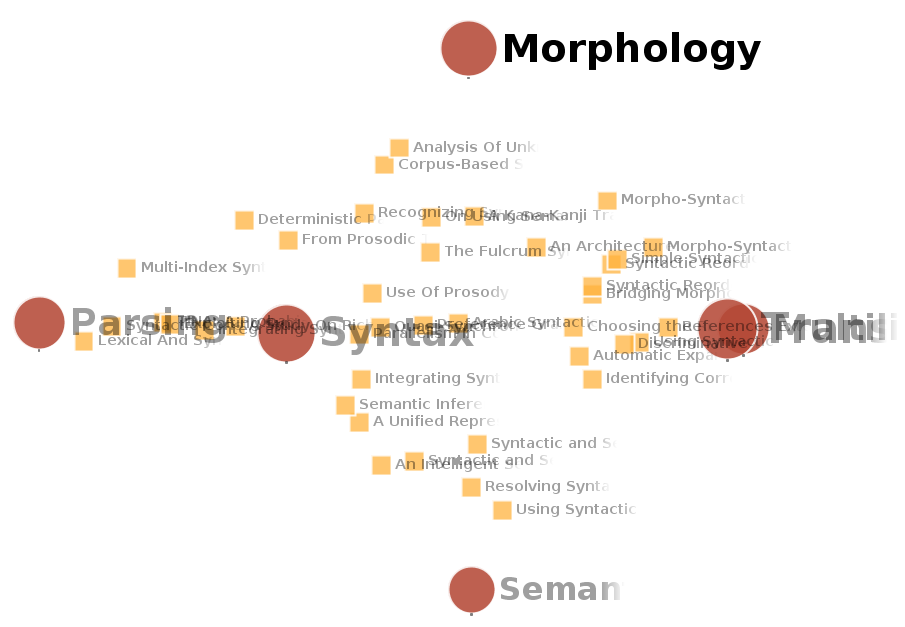}
\caption{The expert user arranges the topic centers on the X and
  Y-axes to create a custom two-dimensional projection of the topic
  space.}
\label{f:regression}
\end{center}
\end{figure}

Suppose that the expert user wants documents that describe
\topic{multilingual} analysis and \topic{translation}, but avoid
\topic{syntax} and \topic{parsing}; in fact, let us suppose that
parsing is completely inappropriate due to technical constraints.  The
user can arrange the topic centers on a line, with \topic{parsing} to
the far left and \topic{syntax} slightly left of center, while
locating the \topic{multilingual} and \topic{translation} topics to
the far right.  Such a configuration can be viewed as a
one-dimensional projection that assigns a large negative weight to
\topic{parsing}, a smaller negative weight to \topic{syntax}, and a
equal positive weights to \topic{multilingual} and
\topic{translation}.  Next, the user wants to distinguish documents
that focus on \topic{morphology} from those that focus on
\topic{semantics} -- this time using the Y-axis.  The final
configuration is shown in Figure~\ref{f:regression}.  The user can now
select the documents in the desired subspace for further viewing and
refinement, as described in the novice scenario.

Overall, we see that in two dimensions, the location of each topic
center defines a projection matrix that reduces the high-dimensional
topic proportion vector to an easily viewable two dimensional
representation.  By dragging topics further from the center, their
absolute weight is increased, causing them to exert a greater
influence on the position of each document.  Thus,
\topicbrowser~offers an intuitive direct manipulation interface for
designing projections that isolate the desired region of topic space.

%% file: system.tex
\section{The TopicViz System}
We now describe in more detail the mechanisms and affordances
underlying the \topicbrowser~system.  The core idea of
\topicbrowser~is to provide affordances for interactively exploring
the topical affiliations of a set of documents, while facilitating
refinement and expansions of the document set.  Thus, the main entry
point is the search query, which will be familiar to users from
traditional information search interfaces.  However, from this point,
we diverge from prior approaches, emphasizing the direct manipulation
design of novel 2D projections and interactive exploration of
document-topic and topic-topic relationships.  The previous section
described the envisioned use cases for such interactions; we now
describe the underlying mechanisms.

\subsection{Document positioning}
As described in Section~\ref{s:background}, a topic model is defined
by the topic-term relations and the topic-document relations.  Both
objects are high-dimensional: the topic-term matrix contains a row for
each topic, and a column for each word in the vocabulary; the
topic-document matrix contains a row for each document and a column
for each topic.  The number of documents and vocabulary size are each
typically in the thousands;\footnote{In the scenario, the number of
  topics is 25; the vocabulary size is 18,743 (after pruning
  infrequent words) the number of documents is 15,032.}, and the
numerical values are not intuitively meaningful on their own; rather,
the structure of the topic model is best understood in a relational
setting.  Thus, we present a document field view incorporating topics
and documents.

In the document field, we see the relationship between documents and
topics.  Inspired by ``dust-and-magnet'' approaches to information
visualization, (e.g., ~\cite{Yi2005Dust}), we initially arrange a ring
of topic nodes around the outside of the field, which act as magnets.
Documents are represented as nodes within this field; edge weights are
based on Hooke's law, with $1-\theta_{di}$ as the force of the spring
between topic $i$ and document $d$.  A document that is a near 100\%
match for a given topic will be placed almost directly on that topic's
magnet; a document that is a 50\% match for each of two topics will be
positioned halfway between them.  Thus, documents that have similar
topic proportions are located near each other, reflecting semantic
differences directly in the spatial layout.  Visualizing such a
high-dimensional model in 2D inevitably causes information to be lost,
but the force-directed layout permits interactive manipulation of
document nodes, allowing users to more closely examine regions of
particular interest.

\subsection{Document set refinement}
As the number of documents in a collection is typically in the
thousands, it is not helpful to view all of the documents at the same
time.  Our interface includes two affordances for selecting sets of
documents to visualize.  The first affordance -- which is the entry
point to interaction with our system -- is the search query.  Just as
in traditional search interfaces, the user enters a query and receives
a list of results (in a separate panel).  These results can be sorted
by traditional metadata: titles, author, year, and venue.  The user
can then drag documents into the document field, which provides an
intuitive graphical visualization of the semantic structure of the
search results.  The second affordance permits the user to walk to the
citation graph, adding citing or cited documents for any set of
documents already in the view; the citation links are made visible.

By default, the set of topic magnets is dynamically updated to show
the topics that are most relevant to the documents currently in the
field.  This feature can be turned off, allowing topics to be added
and deleted manually.

\subsection{Implementation}
\topicbrowser~is implemented through \shiftr~\cite{Shiftr2009Chi}, a
Java platform designed to support interactive exploration and querying
of large graph data with millions of nodes and edges. \shiftr~builds
on the Prefuse library for force-directed
layouts~\cite{Heer2005Prefuse}, providing a collection of fundamental
operations over graph data: querying nodes by arbitrary node
attributes; visualizing user-specified subgraphs; and flexible
spatial arrangement for nodes through pinning and unpinning.
\topicbrowser~uses these lower-level operations to provide a
force-directed layout interface for exploring topic models of document
content.

%% file: summary.tex

\section{Future work}
A key target for future work is empirical validation.  Indeed, beyond
the necessary task of evaluating the specific design decisions taken
in \topicbrowser, we also believe that this tool can serve as a platform 
for 
\emph{in situ} user studies of whether and how topic models can best
support document set exploration and sensemaking.  Specifically, we
plan to develop a battery of information-exploration tasks (similar to
the email exploration tasks of Liu et al.~\cite{Liu2009Interactive})
and compare the efficacy of \topicbrowser~with traditional search
interfaces, as well as textual and table-based representations of
topic models.

From a visualization standpoint, we see several intriguing directions
for future work.  While \topicbrowser~offers an innovative take on the
document-topic relationship, the connection between topics and terms
is still expressed through traditional term lists.  We plan to explore
whether a more spatial visualization for this relationship would be
possible, or whether an alternative approach such as
DocuBurst~\cite{Collins2009DocuBurst} could be incorporated in the
\topicbrowser~environment.  We also believe that an integrated
presentation of document metadata such as time, authorship, and venue
would substantially improve the practical usability of the system.
Finally, we are eager to investigate the use of color as a third
dimension, either to visualize such metadata, or to enable gestalt
high-level comparisons between document
sets~\cite{Landauer2004Paragraph}.

\section{Summary}
Topic models can give powerful insights on document collections -- but
only if used in combination with a comprehensible presentation and an
interaction design built around the information exploration process.
\topicbrowser~presents an interactive visualization that places topic
models in the context of a search interface, filling the same role
currently played by keyword search.  We see two main advantages of our
approach: it accounts for latent document semantics, and provides an
interactive spatial visualization that allows the user to rapidly
focus on key areas of interest.